\documentclass[sigconf]{aamas}

\usepackage{balance} 

\setcopyright{none}
\acmConference[ALA '26]%
{Proc.\@ of the Adaptive and Learning Agents Workshop (ALA 2026)}%
{May 25 -- 26, 2026}%
{Paphos, Cyprus, https://alaworkshop2026.github.io/}%
{Aydeniz, Delgrange, Mohammedalamen, Yang (eds.)}%
\copyrightyear{2026}
\acmYear{2026}
\acmDOI{}
\acmPrice{}
\acmISBN{}
\acmSubmissionID{20}

\title[InfoChess]{InfoChess: A Game of Adversarial Inference and a Laboratory for Quantifiable Information Control}

\author{Kieran A. Murphy}
\affiliation{
  \institution{New Jersey Institute of Technology}
  \city{Newark, NJ}
  \country{United States}}
\email{kieran.murphy@njit.edu}

\begin{abstract}
We propose \textit{InfoChess}, a symmetric adversarial game that elevates competitive information acquisition to the primary objective. 
There is no piece capture, removing material incentives that would otherwise confound the role of information.
Instead, pieces are used to alter visibility.
Players are scored on their probabilistic inference of the opponent’s king location over the duration of the game.
To explore the space of strategies for playing InfoChess, we introduce a hierarchy of heuristic agents defined by increasing levels of opponent modeling, and train a reinforcement learning agent that outperforms these baselines. 
Leveraging the discrete structure of the game, we analyze gameplay through natural information-theoretic characterizations that include belief entropy, oracle cross entropy, and predictive log score under the action-induced observation channel. 
These measures disentangle epistemic uncertainty, calibration mismatch, and uncertainty induced by adversarial movement.
The design of InfoChess renders it a testbed for studying multi-agent inference under partial observability.
We release code for the environment and agents, and a public interface to encourage further study: \texttt{\href{https://github.com/murphyka/infochess}{https://github.com/murphyka/infochess}}.
\end{abstract}


\begin{document}


\pagestyle{fancy}
\fancyhead{}

\maketitle 


\section{Introduction}
Games have long served as laboratories for studying complex phenomena.
While existing games carry historical intuition and strategic depth, novel game design offers the opportunity to isolate specific mechanisms for controlled investigation.
In this work, we introduce a new game---\textit{InfoChess}---as a testbed that elevates competitive information acquisition into its primary objective.

Information is generally a central resource in partially observable games, including poker~\cite{brown2018poker,brown2019multipoker}, Hanabi~\cite{bard2020hanabi}, Stratego~\cite{perolat2022stratego}, many video games~\cite{vinyals2019starcraft,berner2019dota,gilmour2021approach}, and more broadly in partially observable decision-making settings such as POMDPs~\cite{kaelbling1998planning}, where agents must act based on beliefs over latent state.
Partial-information variants of chess have long been explored, such as Kriegspiel~\cite{pritchard2000kriegspiel} and dark chess (fog-of-war chess)~\cite{darkchess,zhang2025FoWchess}.
In these games, however, information is typically instrumental to another goal (e.g., material gain or victory conditions).
As a result, player actions often reflect an entangled mixture of objectives, making it difficult to directly study information acquisition and concealment in isolation.

For example, specific moves of DeepMind’s Stratego agent~\cite{perolat2022stratego} can be interpreted as purposeful deceit (bluffing) when they align with behavior we recognize from human play, but other moves plausibly balance material advantage against information gain or concealment in a combination that is difficult to quantify.
InfoChess instead makes information the explicit and sole objective: players score through continued inference of the opponent king’s location.
This design centralizes information competition rather than embedding it as a secondary effect.
InfoChess thus provides a controlled setting for studying belief modeling, exploration, and strategic concealment in multi-agent partially observable environments.

\begin{figure*}[h]
  \centering
  \includegraphics[width=\textwidth]{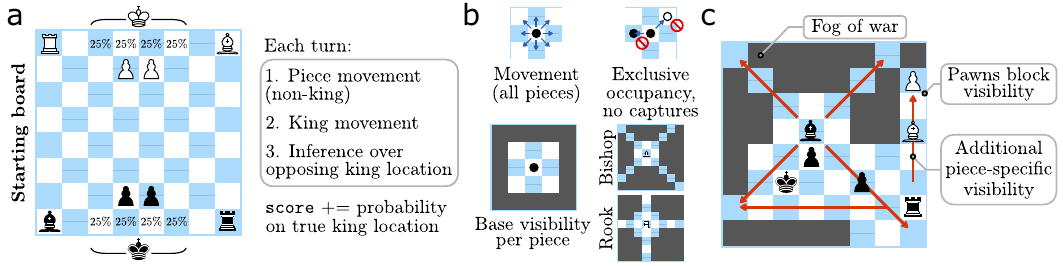}
  \caption{\textbf{The mechanics of InfoChess.} \textbf{(a)} The starting configuration on an $8\times8$ board, where the king is placed randomly among the four indicated squares on each side.
  A turn consists of a non-king piece movement, a king movement, and then an inference about the location of the opponent's king. Probability is assigned to all squares on the board, and the oracle records a score equal to the probability mass placed on the correct (ground truth) square where the opponent's king is located.
  \textbf{(b)} Movement is the same for all pieces: one square in any direction, respecting exclusive occupancy. 
  There are no piece captures in InfoChess.
  Visibility is piece-dependent: all pieces have visibility over their immediate vicinity, but the bishop and rook have viewing rays that match their movement in standard chess.
  \textbf{(c)} An example board state with annotations. 
  Note that pawns obstruct the opponent's vision: viewing rays end at a pawn's square.}
  \label{fig:game}
  \Description{The mechanics of InfoChess}
\end{figure*}

InfoChess can be understood by contrast with standard chess (Fig.~\ref{fig:game}a).
Instead of capturing the opponent king, players aim to infer its location repeatedly throughout the game.
The objective is \textit{adversarial inference}: each player seeks to acquire information about the opponent while minimizing information exposed about their own state.
This framing contrasts with asymmetric adversarial settings such as OpenAI’s multi-agent hide-and-seek environment~\cite{baker2020hideseek}, in which hiders and seekers occupy distinct roles and objectives. 
While hide-and-seek also exhibits emergent information-seeking behavior, the asymmetry of roles complicates interpretation of strategic incentives. 
InfoChess instead presents a symmetric adversarial inference game: both players share identical capabilities and objectives, and each simultaneously acts as seeker and concealer.
There are no piece captures---an intentional design choice that removes material incentives and further foregrounds information dynamics.
All pieces move one square in any direction (Fig.~\ref{fig:game}b), but differ in their visibility effects.
Each piece reveals its immediate surroundings, ensuring legality of movement.
Rooks and bishops cast extended lines of sight analogous to their standard chess motion, while pawns obstruct opponent visibility.
An example board state from Black’s perspective is shown in Fig.~\ref{fig:game}c.

In this work, we focus on three questions: (i) how opponent modeling influences competitive performance, (ii) whether entropy-based characterizations reveal strategic regimes of play, and (iii) whether RL discovers strategies beyond heuristic belief maximization.
In the remainder of this paper, we demonstrate the richness of InfoChess as a laboratory for adversarial information dynamics.
We introduce a hierarchy of heuristic agents that vary in their modeling of hidden state, and train a reinforcement learning (RL) agent that surpasses these baselines.
Using these agents as objects of study, we develop several information-theoretic characterizations that illuminate the strategic structure of the game.
Our code and environment: \texttt{\href{https://github.com/murphyka/infochess}{https://github.com/murphyka/infochess}}.

\section{Game specifics and design rationale}
Figure~\ref{fig:game} illustrates the core mechanics of InfoChess. 
Here we clarify additional rules and the motivations underlying key design choices.
Our design choices aim to isolate adversarial inference while preserving strategic richness and human playability.

\noindent \textbf{Game duration.}
In the absence of piece capture, it becomes natural to terminate games via a fixed horizon.
We use 25 turns per side.
Empirically, a visibility-maximizing agent requires roughly 10–15 turns to reach a plateau in entropy reduction.
To avoid overemphasizing either the initial information-acquisition transient or the subsequent concealment-oriented steady state, we selected a horizon that balances both regimes.

\noindent \textbf{Scoring rule.}
At each turn, a player scores the probability mass assigned to the true opponent king location.
While log-probability would directly reward calibrated uncertainty reduction, linear probability preserves an equivalence in expectation between diffuse and one-hot belief assignments.
This choice simplifies human play by removing the need to explicitly distribute probability mass across all squares at every turn, while maintaining proper incentives for accurate inference.

\noindent \textbf{Initial board configuration.}  
The opponent king is initialized uniformly across four back-row squares.
This restricts extreme opening variability while mitigating first-move advantage.
The remaining piece composition reflects a trade-off: increasing the number of pieces expands action space and strategic depth, but reduces the informational tension by saturating visibility.
For an $8 \times 8$ board, the configuration in Fig.~\ref{fig:game}a strikes a practical balance.
We anticipate that larger boards with more diverse piece sets will yield richer dynamics.

\noindent \textbf{Two movements per turn.}
Allowing only a single movement per turn led to limited king mobility in human play and hindered the design of competitive heuristic agents.
We therefore separate king and non-king movements within each turn.
This decoupling encourages active concealment strategies while preserving offensive visibility maneuvers.

\section{Definitions}
Information gain is a standard concept from active learning used to compare actions in terms of the expected reduction in uncertainty after observation~\cite{houlsby2011bayesian,smith2023prediction}.
Let $q(k|h_{:i}^\alpha)$ denote player $\alpha$'s belief distribution over king locations $k$ given its observed board state history $h_{:i}^\alpha$ up to turn $i$.
For a candidate action $a$, let $\mathcal{F}_a$ denote the set of squares that will be fogged after the move (with the complement being visible).
If the king becomes visible, the posterior collapses to a delta function and has zero entropy (uncertainty).
The only nonzero contribution to the expected posterior entropy therefore arises when the king remains in the fog, in which case we assume the posterior preserves the same relative probabilities over the occluded squares (i.e., the posterior simply renormalizes).

Let
\begin{equation}
    q_f := \sum_{x \in \mathcal{F}_a} q(k=x|h_{:i}^\alpha)
\end{equation} 
be the total probability mass remaining in the fog, and let Shannon entropy be defined as $H(p) = -\sum_i p_i \log p_i$~\cite{cover1999elements}.
Then the expected posterior entropy after action $a$ is
\begin{equation}
    \mathbb{E}[H_\text{post}|a] = q_f H \left( \left \{ \frac{q(k=x|h_{:i}^\alpha)}{q_f} \right \} _{x \in \mathcal{F}_a} \right),
\end{equation}
and the expected information gain is $\Delta H_a = H_\text{prior} - \mathbb{E}[H_\text{post}|a]$.
While information gain guides action selection, we introduce additional entropy-based quantities in Section~\ref{sec:characterization} to characterize emergent gameplay dynamics.

\section{Agents: Policies and beliefs}

We define a hierarchy of heuristic agents that differ in their degree of opponent modeling, followed by a learned RL agent.

\subsection{Belief Models}

If the opponent king is visible, all agents assign unit probability to that location. 
Otherwise, two belief models are considered.

\textbf{Uniform.}  
Probability mass is distributed uniformly across all fogged squares, regardless of past observations (e.g., if the king was observed and then moved into fog, all fogged squares are assigned uniform probability regardless of distance).

\textbf{Learned.}  
A two-layer transformer (four heads per layer, hidden dimension 128) processes the history of board states from the player’s perspective. 
The resulting representation is fed to an MLP head that outputs a distribution over board squares.
Training via cross-entropy loss uses ground-truth opponent king locations.
For input to the transformer, board states are canonicalized to the acting player's perspective (i.e., transformed as if the player is always white) and encoded as multi-channel tensors indicating team identity, piece identity, and square visibility for a flattened shape of $8\times8\times6=384$.
Training was for 15 epochs using Adam and a learning rate of $10^{-3}$.

A second locus of uncertainty concerns whether the player’s own king is visible to the opponent. 
We model this analogously:

\textbf{Uniform visibility.}  
All legal king squares are treated as equally likely to be visible.

\textbf{Learned visibility.}  
A second MLP head atop the shared transformer trunk (described above) predicts per-square opponent visibility and is trained with binary cross-entropy. 
The king-belief and visibility heads are trained jointly via a summed loss.

Training data for both heads consists of 10,000 games generated from random mixtures of Random and VisMax agents (defined below).

\begin{table*}[t]
\centering
\begin{tabular}{lllll}
\hline
\textbf{Agent Name} & \textbf{Piece policy} & \textbf{King policy} & \textbf{King belief} & \textbf{Opponent visibility belief} \\
\hline
Random & Random & Random & Uniform & Uniform \\
VisMax (V) & Greedy info gain & Random & Uniform & Uniform \\
BeliefMax (B) & Greedy info gain & Random & Learned & Uniform \\
HidingVisMax (HV) & Greedy info gain & Greedy concealment & Uniform & Learned \\
HidingBeliefMax (HB) & Greedy info gain & Greedy concealment & Learned & Learned \\
RL Agent & Learned & Learned & Learned & Learned \\
\hline
\end{tabular}
\caption{\textbf{Agent definitions in terms of movement policies and belief models.}
King and opponent visibility belief models are trained with privileged full-state supervision but operate on partial observations at evaluation.}
\label{tab:agent_comparison}
\end{table*}

\subsection{Heuristic Agents}

Agents are defined by movement policies and belief modeling (Tab.~\ref{tab:agent_comparison}).

\begin{enumerate}
    \item \textbf{Random} selects all moves uniformly at random and uses the uniform king belief.
    
    \item \textbf{VisMax (V)} greedily maximizes newly visible squares with non-king moves (equivalently, information gain under a uniform belief). King movement is random.
    
    \item \textbf{BeliefMax (B)} greedily maximizes expected information gain under the learned king belief model. King movement is random.
    
    \item \textbf{HidingVisMax (HV)} follows the VisMax non-king policy. The king moves to the legal square with minimal predicted opponent visibility.
    
    \item \textbf{HidingBeliefMax (HB)} combines BeliefMax non-king movement with the HidingVisMax king policy.
\end{enumerate}

\subsection{Reinforcement Learning Agent}

We additionally train an RL agent using the shared transformer trunk as a frozen state encoder. 
The trunk is frozen during RL to isolate the effect of policy learning from representation learning.
For each candidate move, the trunk representation is concatenated with a movement vector (8 dimensions: $(x_0, y_0, x_1, y_1)$ and a one-hot vector indicating the piece identity) and passed to an MLP that outputs a scalar score.
Note the RL agent uses the same MLP to score both the king and non-king moves, and the inference step uses the same learned king belief model as the BeliefMax agent.

The scoring network is trained with REINFORCE~\cite{williams1992REINFORCE} to maximize per-turn score differentials. 
Training is performed over 45,000 episodes (batch size 10 games) against a mixture of opponents:
30\% self-play,
5\% Random,
15\% VisMax,
15\% BeliefMax,
15\% HidingVisMax,
20\% HidingBeliefMax.

Optimization uses Adam with learning rate $3\times10^{-4}$. 
The objective includes an entropy bonus equal to the sum of the entropies of the non-king and king move distributions, with coefficient linearly annealed from $5\times10^{-2}$ to $5\times10^{-3}$ over training.

\section{Characterizing game dynamics}
\label{sec:characterization}
\begin{figure*}[h]
  \centering
  \includegraphics[width=\linewidth]{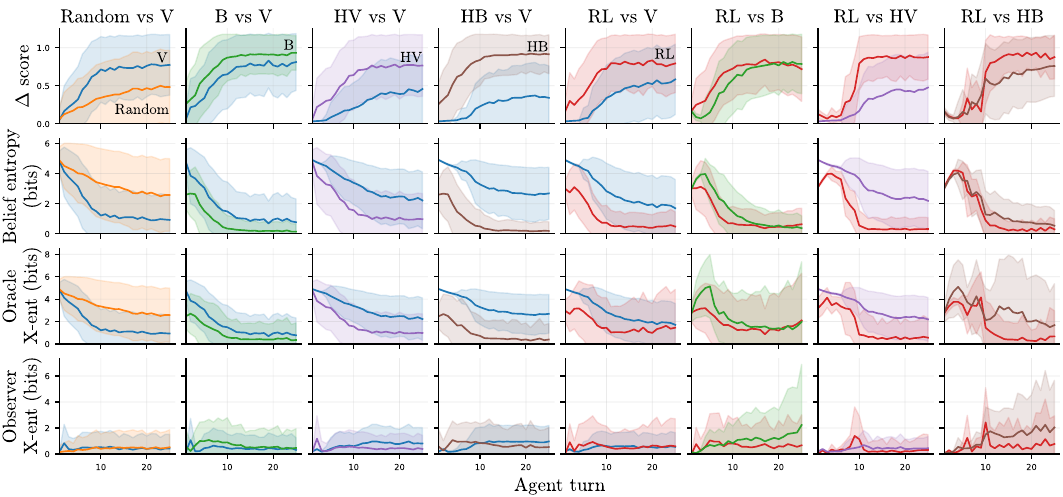}
  \caption{\textbf{Per-turn characterization.} The curves show the mean of the quantity, averaged per turn per agent over 250 games per matchup, and the shaded regions indicate one standard deviation in either direction.
  ``$\Delta$ score'' refers to the change in score earned during that turn.}
  \label{fig:avg_metrics}
  \Description{Per-turn characterization of all matchups against the VisMax and RL agents.}
\end{figure*}

The central role of information in InfoChess, together with its discrete state space, facilitates several information-theoretic characterizations of the partial-information states encountered by players. 
While oracle metrics access the hidden state directly, the agent itself observes the environment only through an action-dependent observation channel, which limits the identifiability of its belief over the hidden state.
For selected pairwise matchups between agents, we track per-turn score along with the quantities defined below as a function of turn (Fig.~\ref{fig:avg_metrics}). 
Curves indicate the mean over 100 games, with shaded regions showing a standard deviation in either direction. 
Games were played with a random split of white and black assignments.

\subsection{Belief entropy}
A first-order characterization of a player’s uncertainty is the entropy of its belief distribution over the opponent king location (for brevity, we write $q(k)$ for $q(k|h_{:i}^\alpha)$):
\begin{equation}
    H(q) = - \sum_k q(k) \log q(k).
\end{equation}
This quantity measures total epistemic uncertainty about the latent state.

For the Random and VisMax agents, the belief is uniform over all fogged squares, and therefore the entropy reduces to the logarithm of the number of occluded squares (i.e., the area of the fog of war). 
More sophisticated agents may maintain non-uniform beliefs reflecting inferred structure.

\begin{figure}[h]
  \centering
  \includegraphics[width=\linewidth]{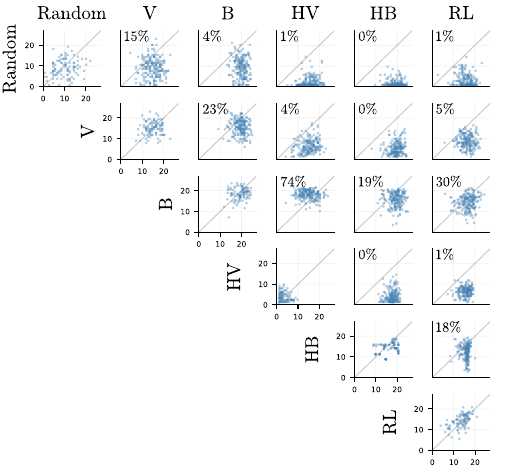}
  \caption{\textbf{Pairwise score distributions.} Each scatter plot displays the final scores for 100 self-play matches (diagonal) or 200 cross-agent matches (100 for black-white and 100 for white-black).
  The horizontal axis represents the score for the agent corresponding to the label at the top of the column, and the percentage reflects the proportion of games won by the agent listed to the left of each row (whose score is on the vertical axis).}
  \label{fig:matchup}
  \Description{Pairwise score distributions}
\end{figure}

\subsection{Oracle cross entropy}
Belief entropy measures uncertainty but does not measure correctness. 
To assess calibration relative to the true latent state, we compute the oracle cross entropy of the agent’s belief $q(k)$ with respect to the true distribution $p(k)$ over king locations:
\begin{equation}
    H(p,q) = \mathbb{E}_{k \sim p(k)}[-\log q(k)] 
    = H(p) + D_{\mathrm{KL}}(p(k)\,\|\,q(k)).
\end{equation}

This quantity decomposes into two components: 
the entropy $H(p)$ of $p(k)$, the true distribution over king locations induced by marginalizing over unobserved variables (e.g., opponent behavior and hidden state), and the KL divergence measuring mismatch between belief and truth. 
When $q(k) = p(k)$, the KL term vanishes and the oracle cross entropy equals the entropy of the opponent-induced state distribution.

Operationally, this may be interpreted as the expected surprisal the agent would experience if the king location were revealed on every turn.

\subsection{Observer cross entropy}

In practice, the ground-truth king location is not revealed each turn. 
Instead, the agent observes only whether the king becomes visible (and at which square), or remains hidden. 
This induces a coarsened observation channel determined by the chosen action.

Let $q(k)$ denote the agent’s belief before taking action $a$. 
The action partitions board squares into visible squares $V_a$ and fogged squares $\mathcal{F}_a$. 
The resulting observation variable $O_a$ takes values
\[
    O_a =
    \begin{cases}
        k, & k \in V_a \quad (\text{king revealed at }k), \\
        \text{hidden}, & \text{if } k \in \mathcal{F}_a.
    \end{cases}
\]

This defines a pushforward distribution over observations:
\[
    q_O(o) = 
    \begin{cases}
        q(k=o), & o \in V_a, \\
        \sum_{x \in \mathcal{F}_a} q(k=x), & o = \text{hidden}.
    \end{cases}
\]

Let $p_O(o)$ denote the corresponding true observation distribution induced by the opponent’s policy. 
We define the observer cross entropy as
\begin{equation}
    H(p_O, q_O) = \mathbb{E}_{o \sim p_O}[-\log q_O(o)].
\end{equation}

While our agents are trained with privileged access to the hidden state, it is instructive to consider the learning problem under partial observability alone.
The observer cross entropy corresponds to a strictly proper scoring rule over the observation space, ensuring that an agent trained solely from partial observations would converge to $p_O$~\cite{gneiting2007strictly}. 
However, because learning proceeds through the coarsened observation channel, this guarantees recovery only of the projected distribution 
$p_O$, not the latent state distribution ($p(k)$).

Observer cross entropy evaluates predictive calibration through the action-induced observation channel. 
Because $O_a$ is a deterministic coarsening of the latent king state, the data processing inequality implies
\[
    D_{\mathrm{KL}}(p_O \,\|\, q_O) 
    \le 
    D_{\mathrm{KL}}(p(k) \,\|\, q(k)),
\]
so detectable mismatch after projection can only decrease relative to the latent distribution.

When the agent’s belief matches the true king distribution, the KL term vanishes and the observer cross entropy reduces to the entropy of the induced observation distribution  $H(p_O)$. 
This residual reflects irreducible uncertainty introduced by the observation channel rather than model miscalibration. 
Thus, cross entropy captures expected surprisal under partial observability, while $D_{\mathrm{KL}}(p_O \,\|\, q_O)$ isolates miscalibration.

\begin{figure}[h]
  \centering
  \includegraphics[width=0.9\linewidth]{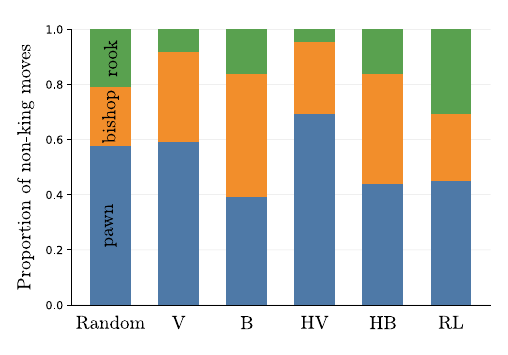}
  \caption{\textbf{Non-king movement by agent.} For each agent in 1,000 matches against randomly selected opponents (including self-play), the allocation of non-king movement is shown across the three types of pieces.}
  \label{fig:movement}
  \Description{Breakdown of non-king movement.}
\end{figure}

\section{Discussion}
With only two levels of opponent modeling, the heuristic agents (Tab.~\ref{tab:agent_comparison}) exhibit a hierarchy of competitiveness.
The results of hundreds of matches (Fig.~\ref{fig:matchup}) show how greedily maximizing information gain about the opponent king location  with respect to a learned belief model increases the average score earned by the player (V$\rightarrow$B, HV$\rightarrow$HB).
Orthogonally, greedily moving the king based on a learned belief model of the opponent's visibility reduces the opponent's score (V$\rightarrow$HV, B$\rightarrow$HB).
The strongest heuristic agent leverages both belief models to guide piece movement (Hiding BeliefMax, HB).

The RL agent outperforms all the heuristic agents.
Interestingly, the BeliefMax agent is the strongest competitor against it.
Presumably the RL agent found a way to exploit the reduced stochasticity of hiding behavior; VisMax slightly outperforms the Hiding VisMax against RL as well.

By looking at per-turn characterizations in different matchups (Fig.~\ref{fig:avg_metrics}), hiding behavior clearly drops the score and increases the entropy of the belief distribution over the opponent king location against the VisMax agent.
The Hiding BeliefMax agent exhibits both benefits again: higher per turn scores from the informed belief model over the opposing king location, and higher opponent entropy.
The RL agent appears to outperform the BeliefMax agent only in the initial turns of the game, and outperforms the Hiding BeliefMax agent in the later part of the game.
The oracle cross entropy largely tracks the actual entropy of the belief distribution, suggesting the latter is well calibrated.
Deviations can be seen in matches against the RL agent.
The observer-based cross entropy shows relatively little signal for the weaker agents, but indicates mismatch in the later stages of the game against the RL agent, which suggests the RL agent moves against what the belief model predicts.

A further window into the different strategies is the allocation of non-king moves by piece type (Fig.~\ref{fig:movement}).
We show the fraction of non-king moves across 1,000 matches for each agent, played against an even split of all possible opponents.
Without an informed belief model for the opponent king, the VisMax agents move pawns even more frequently than Random, as pawns reveal several squares close to the player's own side of the board. 
In contrast, the pawn proportion drops considerably for the BeliefMax and RL agents.
Interestingly, while the bishop is the next most moved piece for all heuristic agents, the RL agent moves the rook more often than the bishop.

InfoChess isolates adversarial inference as a first-class objective. 
Our results show that modest opponent modeling already yields clear gains, and that information-theoretic metrics expose distinct regimes of play. 
More broadly, the game highlights how action-dependent observation channels constrain what can be learned about latent state. 
This makes InfoChess a useful testbed for studying inference, concealment, and coordination in partially observable multi-agent systems.





\balance



\bibliographystyle{ACM-Reference-Format} 
\bibliography{references}

\end{document}